\input{aipcheck}

\documentclass[
    ,final            
  ]
  {aipproc}

\layoutstyle{6x9}

\begin{document}

\title{The NASA/IPAC/NExScI Star And Exoplanet Database}


\author{G. B. Berriman}{
  address={NASA Exoplanet Science Institute and Infrared Processing and Analysis Center, California Institute of Technology}
}

\author{ B. Ali}{
 address={Infrared Processing and Analysis Center, California Institute of Technology} 
}

\author{R. Baker}{
  address={Infrared Processing and Analysis Center, California Institute of Technology}
}

\author{K. Von Braun}{
  address={NASA Exoplanet Science Institute, California Institute of Technology}
}

\author{N-M. Chiu}{
  address={Infrared Processing and Analysis Center, California Institute of Technology}
}

\author{D. R. Ciardi}{
  address={NASA Exoplanet Science Institute, California Institute of Technology}
}

\author{J. Good}{
  address={Infrared Processing and Analysis Center, California Institute of Technology}
}
\author{S. R. Kane}{
  address={NASA Exoplanet Science Institute, California Institute of Technology}
}


\author{M. Kong}{
  address={NASA Exoplanet Science Institute, California Institute of Technology}
}

\author{A. C. Laity}{
  address={Infrared Processing and Analysis Center, California Institute of Technology}
}

\author{ D. L. McElroy}{
  address={Infrared Processing and Analysis Center, California Institute of Technology} 
}

\author{ S. Monkewitz}{
  address={Infrared Processing and Analysis Center, California Institute of Technology}
}

\author{ A. N. Payne}{
  address={NASA Exoplanet Science Institute, California Institute of Technology}
}

\author{S. Ramirez}{
  address={NASA Exoplanet Science Institute, California Institute of Technology}
}

\author{ M. Schmitz}{
  address={Infrared Processing and Analysis Center, California Institute of Technology}
}

\author{J. S. Stauffer}{
  address={Spitzer Science Center, California Institute of Technology}
}

\author{P. L. Wyatt}{
  address={NASA Exoplanet Science Institute, California Institute of Technology}
}

\begin{abstract}
 
 The NASA/IPAC/NExScI Star and Exoplanet Database (NStED) is a general purpose stellar archive which supports NASA planet-finding and planet-characterization goals, stellar astrophysics, and the planning of NASA and other space missions. There are two principal components of NStED: a database of 140,000 nearby stars and exoplanet-hosting stars, and an archive dedicated to high precision photometric surveys for transiting exoplanets (NStED-ETSS). We present summaries of these components. The NStED stellar database currently serves published parameters for 140,000 stars. These parameters include coordinates, multiplicity, proper motion, parallax, spectral type, multiband photometry, radial velocity, metallicity, chromospheric and coronal activity index, rotation velocity/period, infrared excess. NStED-ETSS currently serves data from the TrES survey of the Kepler field as well as dedicated photometric surveys of four stellar clusters. NStED-ETSS aims to serve both the surveys and the broader astronomical community by archiving these data and making them available in a homogeneous format.
 
\end{abstract}

\maketitle


\section{Introduction}

The NASA Star and Exoplanet Database (NStED)  is dedicated to
collecting and serving vital published data involved in the search
for and study of extrasolar planets and their host stars. 

NStED consists of two sets of services:
\begin{itemize}
\item The stellar
and exoplanet services provide access to stellar parameters of
potential exoplanet bearing stars along with exoplanet parameters, and
\item The Exoplanet Transit Survey Service (ETSS) provides an interface dedicated to searches of exoplanet transit surveys.
\end{itemize}

Currently, all these services are accessible through simple web forms at at the project web page \footnote{http://nsted.ipac.caltech.edu}.  The following sections describe the important features of the stellar and ETSS services. 

\section{ Stellar Services In NStED}
The
stellar services provided by NStED include the following:
\begin{itemize}
\item Access to data related to relatively bright nearby stars.
\item The capability to display and visualize the properties of individual stars. 
\item The capability to perform complex searches on stellar and planetary parameters.
\item Access to published images, spectra, and time series data related to the
  stars in the database.
\end{itemize}
Complementary to this are the exoplanet services, which include the
following:
\begin{itemize}
\item General data and pubished parameters for the known exoplanets
  and host stars.
\item Photometric and radial velocity data related to the known
  exoplanets.
\end{itemize}

\section{Stellar Content in  NStED}

NStED's stellar and exoplanet content is composed of published tabular
data, derived and calculated quantities, and associated data including
images, spectra, and time series. Some of data sets have been generously contributed by data providers, such as echelle spectra  from the N2K
consortium (\cite{fis05}).

NStED's core set of stars is derived from the Hipparcos,
Gliese-Jahreiss, and Washington Double Star catalogs. The total number
of Hipparcos and Gliese-Jahreiss stars within NStED is approximately
140,000. A summary of the stellar parameters and data within NStED is
shown in Table 1. NStED currently supports complex multi-faceted
queries on approximately 75 astrophysical stellar and exoplanet
parameters.

\begin{table}
    \caption{Summary of stellar content within NStED.}
    \begin{tabular}{@{}lll}
      \hline
      Published Parameters & Derived Parameters & Associated Data \\
      \hline
      Position, Distances & Temperature      & Images  \\
      Kinematics          & Luminosity       & Spectra \\
      Photometry, Colors  & Radius           & \\
      Spectral Type       & Mass             & \\
      Luminosity Class    & LSR Space Motion & \\
      Metallicity         & & \\
      Rotation            & & \\
      Activity Indicators & & \\
      Variability         & & \\
      Multiplicity        & & \\
      \hline
    \end{tabular}
\end{table}

\section{Exoplanet Content for NStED}

In order to facilitate future exoplanet studies, NStED maintains an
up-to-date list of exoplanetary systems and associated stellar data by
monitoring daily the literature and making weekly updates to the
database.  The
predicted signatures of exoplanets are also calculated to aid users in
selection of stars appropriate for planet searching and
characterization. The exoplanet signature predictions include
habitable zone sizes, astrometric and radial velocity wobbles, and
transit depths. A summary of the exoplanet parameters and data in
NStED is shown in Table 2.


\begin{table}
    \caption{Summary of exoplanet content within NStED.}
    \begin{tabular}{@{}lll}
      \hline
      Published Parameters & Predicted Parameters & Associated Data \\
      \hline
      Number of Planets & Habitable Zone          & High Contrast Images \\
      Planetary Mass    & Astrometric Wobble      & Lightcurves  \\
      Orbital Period    & Radial Velocity Wobble  & \\
      Orbital semi-major axis & Earth V Magnitude & \\
      Orbital Eccentricity & Earth 10 $\mu$m flux density & \\
      Link to entry in the & & \\
      Exoplanet Encyclopaedia & & \\
      \hline
    \end{tabular}
\end{table}

\section{Specific Goals of NStED-ETSS}

The purpose of NStED-ETSS is to make available to the astronomical community
time-series light-curves of planet transit studies and other
variability surveys in a homogeneous format, along with tools for data
analysis and manipulation. The principal goals of NStED-ETSS include the
following:
\begin{itemize}
\item Provide access to support data for ground-based and space-based transit missions.
\item Support optimization of algorithms for transit
detection or variability classification on existing survey data sets;
for instance, to enable the detection of planets previously missed in the
original study.
\item Extend the time baseline for transit studies by using data sets
containing the same stars, leading to increased detection efficiency, results
of increased statistical significance, enhanced potential to conduct transit
timing studies, etc.
\item Enable improved understanding of false positivies encountered in transit
surveys.
\item Provide access to a wealth of other astrophysical results and ancillary
science not pursued in the original survey, such as studies of eclipsing
binary and other variable stars or variability phenomena, stellar atmospheres
(rotation, flares, spots, etc.), asteroseismology and intrinsic stellar
variability, as well as serendipitous discoveries such as photometric
behaviors of supernovae progenitors, etc.
\end{itemize}

\section{ETSS Holdings and Future Data Sets}\label{holdings}

Here we summarize the data sets accessible through ETSS. All the data are organized in a common ASCII format 
for portability:
a master file provides  the basic properties of the data set and
parameters describing the light curves,  one for each star in the data set.

TrES-Lyr1, the TrES network planet transit survey of a field in Lyra,
described in \cite{ocm06}, contains $\sim$ 26,000 stars with 15,500
observation epochs over 75 nights in the $R$ and $r$ filters. The data sets on
the globular clusters (GCs) M10 and M12 contain 44,000 and 32,000 stars,
respectively, with $\sim$ 50 observational epochs in both $V$ and $I$ over a
500-night timespan (\cite{bmc02}). The data set on the GC NGC 3201 features
$\sim$ 59,000 stars with 120 epochs in each $V$ and $I$ over the course of 700
nights (\cite{bm01, bm02}). NGC 2301 is an open cluster and its data set
contains 150 epochs in $R$ on 4,000 stars over 14 nights (\cite{hvt05,the05}).
The KELT-Praesepe data set \cite{ppd07,psp08} contains  light curves of 66,637 stars at $R_K$ with 3,00 epochs over 73 nights 

Data sets will be ingested in 2008 and 2009 include  WASP0 (PI:
S. R. Kane), VULCAN (PI: N. Batalha), BOKS (PIs: S. Howell \&
J. J. Feldmeier), EXPLORE/OC (PIs: K. von Braun \& B. L. Lee), as well as
future CoRoT fields, as NStED is collaborating with the CoRoT team to provide
a NASA portal to the public CoRoT data \cite{b06}.

Each featured data set has been graciously donated by the respective survey
team.  Astronomers wishing to donate data sets are invited to contact the NStED Help Desk \footnote {http://nsted.ipac.caltech.edu/cgi-bin/Helpdesk/nph-genTicketForm}.




\begin{theacknowledgments}
  
  The NASA/IPAC/NExScI Star and Exoplanet Database is operated by the Jet Propulsion Laboratory, California Institute of Technology, under contract with the National Aeronautics and Space Administration.

\end{theacknowledgments}




\bibliography{sample}

\begin{thebibliography}{}

\bibitem{b06}
Baglin, A.\ 2006,
in The CoRoT Mission, \textit{ESA-SP-1306}, 33

\bibitem{hvt05} 
Howell, S.~B., VanOutryve, C., Tonry, J.~L., Everett, M.~E., 
\& Schneider, R.\ 2005, 
\textit{PASP}, 117, 1187 

\bibitem{fis05}
Fischer, D.A., et al.\ 2005,
\textit{ApJ}, 620, 481

\bibitem{ocm06} 
O'Donovan, F.~T., et al.\ 2006, 
\textit{ApJ}, 651, L61 

\bibitem{psp08} 
Pepper, J., Stanek, K.~Z., Pogge, R.~W., Latham, D.~W., DePoy, D.~L., Siverd,
R., Poindexter, S., \& Sivakoff, G.~R.\ 2008, 
\textit{AJ}, 135, 907

\bibitem{ppd07} 
Pepper, J., et al.\ 2007, 
\textit{PASP}, 119, 923

\bibitem{the05} 
Tonry, J.~L., Howell, S.~B., Everett, M.~E., Rodney, S.~A., Willman, M., 
\& VanOutryve, C.\ 2005, 
\textit{PASP}, 117, 281 

\bibitem{bm01} 
von Braun, K., \& Mateo, M.\ 2001, 
\textit{AJ}, 121, 1522

\bibitem{bm02} 
von Braun, K., \& Mateo, M.\ 2002, 
\textit{AJ}, 123, 279

\bibitem{bmc02} 
von Braun, K., Mateo, M., Chiboucas, K., Athey, A., \& Hurley-Keller, D.\
2002, 
\textit{AJ}, 124, 2067 

\end{thebibliography}

\IfFileExists{\jobname.bbl}{}
 {\typeout{}
  \typeout{******************************************}
  \typeout{** Please run "bibtex \jobname" to optain}
  \typeout{** the bibliography and then re-run LaTeX}
  \typeout{** twice to fix the references!}
  \typeout{******************************************}
  \typeout{}
 }


\end{document}